\documentclass[aps,prl,twocolumn,showpacs,superscriptaddress,groupedaddress]{revtex4}  % for review and submission
\usepackage{graphicx}  % needed for figures
\usepackage{dcolumn}   % needed for some tables
\usepackage{bm,amsmath}        % for math
\usepackage{amssymb}   % for math
\usepackage{color}

\newcommand{\beq}{\begin{equation}} 
\newcommand{\eeq}{\end{equation}} 
\newcommand{\bea}{\begin{eqnarray}} 
\newcommand{\eea}{\end{eqnarray}} 

\newcommand{\cev}[1]{\reflectbox{\ensuremath{\vec{\reflectbox{\ensuremath{#1}}}}}}

\begin{document}

%Preprints
\hspace{5.24in} 
\mbox{CERN-PH-TH/2013-061} 
\vspace{0.2in}
\hspace{5.2in} 
\mbox{FERMILAB-PUB-13-063-T}

\title{Geolocating the Higgs Boson Candidate at the LHC}
\author{James~S.~Gainer} \affiliation{Physics Department, University
  of Florida, Gainesville, FL 32611, USA}
\author{Joseph~Lykken} \affiliation{Theoretical Physics Department, Fermilab, Batavia, IL 60510, USA}
\author{Konstantin~T.~Matchev} \affiliation{Physics Department,
  University of Florida, Gainesville, FL 32611, USA}
\author{Stephen Mrenna} \affiliation{SSE Group, Computing Division, Fermilab, Batavia, IL 60510, USA}
\author{Myeonghun~Park} \affiliation{Theory Division, Physics
  Department, CERN, CH-1211 Geneva 23, Switzerland}
\date{July 26, 2013}

\begin{abstract}
The latest results from the ATLAS and CMS experiments 
at the CERN Large Hadron Collider (LHC) unequivocally
confirm the
existence of a resonance, $X$, with mass near $125$ GeV
which could be the Higgs boson of the Standard Model.
Measuring the properties (quantum numbers and couplings) of this 
resonance is of paramount importance. 
Initial analyses by the LHC collaborations
disfavor specific alternative benchmark hypotheses, e.g.~pure pseudoscalars 
or gravitons.  However, this is just the first step in a long-term program 
of detailed measurements.
We consider the most general set of operators in the decay channels
$X\to ZZ,WW,Z\gamma,\gamma\gamma$ and derive the constraint implied
by the measured rate. This allows us to provide a useful
parametrization of the orthogonal independent Higgs coupling degrees
of freedom as coordinates on a suitably defined sphere.
\end{abstract}

\maketitle
{\bf Introduction.} The determination of the spin $J$ and parity $P$ 
of the putative Higgs boson~\cite{discovery} is of paramount importance. 
The observed decay to $\gamma\gamma$~\cite{gammagamma}
already excludes 
the case of $J=1$ by the Landau-Yang theorem~\cite{LY}.
Spin discrimination and coupling measurements 
are achieved by studying kinematic correlations of the final state
objects, generally leptons. An optimal way of incorporating 
all available information is through the Matrix Element Method
(MEM)~\cite{mem gen, jhu mem, mem, mekd}.
Initial results from ATLAS~\cite{ATLAS} and CMS 
~\cite{CMS, CMS-PAS} strongly disfavor $J^P=0^-$, as well as $J^P=2^+$ with
graviton-like couplings.  These results represent an
important first foray into the general parameter space of generic
boson couplings.

In this paper we provide a theoretical framework for 
spin and coupling measurements in full generality,
without theoretical prejudice towards specific benchmarks. 
In contrast to recent global fits of Higgs couplings~\cite{scan,CP},
our approach is geared towards a more model independent measurement in
a single channel, including all relevant operators and utilizing
all available kinematic information. 
For concreteness and simplicity, we consider
the example of a spin-zero ``$X$'' resonance, decaying to four leptons 
through two intermediate $Z$-bosons~\cite{hzz_m2, hzz, impost, ZZ BG}, though our approach 
can be readily extended to other spins and final states. 
The complete measurement of the $X$ couplings in full generality 
will provide insights into the nature of electroweak symmetry breaking
and may offer the first glimpses of physics beyond the Standard Model (SM).

{\bf Effective Theory.}
In what follows, we will consider a spin-zero state $X$.
In general, $X$ has no definite CP properties, and can be thought 
of as a linear combination of some $CP$-even state $H$ and some 
$CP$-odd state $A$~\cite{CP}:
\beq
X \equiv H \cos\alpha + A \sin\alpha.
\label{Xdef}
\eeq
In the special case of $\alpha=0$, $X$ is a pure $J^P=0^+$ scalar, as predicted 
in the Standard Model, while for $\alpha=\pm\pi/2$, $X$ is a pure pseudoscalar with
$J^P=0^-$.

The $X$ couplings to two gauge bosons (e.g.~$ZZ$) 
can be classified according to their symmetries in the $CP$-eigenstate basis $(H,A)$. 
Three types of terms are possible:
\bea
{\cal L} &\supset & -\frac{M_Z^2}{v} H Z^\mu \hat{f}^{(H)}_{\mu\nu} Z^\nu
- \frac{1}{2}H F^{\mu\nu} \hat{f}^{(H)}_{\mu\nu\rho\sigma} F^{\rho\sigma} \nonumber \\
&-& \frac{1}{2}A F^{\mu\nu} \hat{f}^{(A)}_{\mu\nu\rho\sigma} F^{\rho\sigma},
\label{LHA}
\eea
where $F^{\mu\nu}$ is the $Z$-boson field strength tensor,
$M_Z=91.1876$ GeV is the $Z$-boson mass, $v=246$ GeV is the electroweak scale,
and $\hat{f}^{(H)}$ and $\hat{f}^{(A)}$ are (in general 
momentum-dependent) form-factors, which
from the effective theory point of view should be thought of as
infinite series expansions in terms of some new physics scale $\Lambda$.
Let us discuss each one in turn.

The form-factor $\hat{f}^{(H)}_{\mu\nu}$ describes 
interactions of the CP even component $H$ that 
necessarily violate $SU(2)$ gauge invariance.
Expanding in powers of $\Lambda^{-1}$, one obtains
\bea
\hat{f}^{(H)}_{\mu\nu} &\equiv& g_1 g_{\mu\nu}
+ \frac{g_5}{\Lambda^2} \left( \vec\partial_\mu \cev\partial_\nu +
  g_{\mu\nu} \vec\partial^\rho \cev\partial_\rho\right) \nonumber \\ 
&+& \frac{g_6 }{\Lambda^2} g_{\mu\nu} \left( \cev{\Box} + \vec{\Box} \right)+
{\cal O}\left( \frac{1}{\Lambda^4} \right), \label{fH2}
\eea
where $g_i$ are dimensionless coupling constants, and the derivative
operator $\cev{\partial}$ ($\vec{\partial}$) 
acts on the $Z$ field to its left (right).  
Eq.~(\ref{fH2}) does not include terms containing $\partial^\mu
Z_\mu$, which vanish in the Lorenz gauge.

The second class of $CP$-even couplings, described by the form-factor
\beq
\hat{f}^{(H)}_{\mu\nu\rho\sigma} \equiv \frac{g_2}{\Lambda} g_{\mu\rho}g_{\nu\sigma} 
+ \frac{g_3}{\Lambda^3} g_{\mu\rho} \partial_\nu \partial_\sigma + {\cal O}\left( \frac{1}{\Lambda^5} \right),
\label{fH4}
\eeq
may respect $SU(2)$ gauge invariance, if $H$ is a SM singlet.
Note that terms of this sort could be generated by a $\left( \vec\partial_\mu \cev\partial_\nu -
  g_{\mu\nu} \vec\partial^\rho \cev\partial_\rho\right)$ term 
in Eq.~(\ref{fH2}), which explains the absense of such a term there.
Finally, the couplings of the $CP$-odd component $A$ are given by
\beq
\hat{f}^{(A)}_{\mu\nu\rho\sigma} =
\frac{g_4}{2 \Lambda}  \varepsilon_{\mu\nu\rho\sigma}  
 + \mathcal{O}\left(\frac{1}{\Lambda^5}\right)
\label{fA4}
\eeq
and may be $SU(2)$ gauge invariant as well, if $A$ is a SM singlet.
However, here we do not make any assumptions about the $SU(2)$ quantum numbers of
$H$ and $A$; the terms (\ref{fH4}) and (\ref{fA4}) will also violate $SU(2)$ gauge
invariance if $H$ or $A$ (respectively) transform under $SU(2)$.

We note that the $g_5$ and $g_6$ terms in (\ref{fH2}) are typically omitted in the literature.
The contributions to the $X\to ZZ$ amplitude
resulting from the $\cev{\Box} + \vec{\Box} $ and $\vec\partial^\rho \cev\partial_\rho$ terms 
are proportional to $M_{Z_1}^2 + M_{Z_2}^2$ and $M_X^2 - M_{Z_1}^2 - M_{Z_2}^2$,
respectively, where $M_X$ is the mass of the resonance and $M_{Z_1}$ and
$M_{Z_2}$ are the invariant masses of the two $Z$-bosons (in the usual convention where
$M_{Z_1}>M_{Z_2}$).   
Thus, if both $Z$-bosons were on-shell, the contribution from 
such terms would be essentially constant, and therefore could be absorbed
into a redefinition of $g_1$.  
However, in the case of interest where $M_X \approx 125$ GeV,
one or both of the $Z$'s are off-shell and $M_{Z_1}$ and $M_{Z_2}$ vary from
event to event. 
Thus, strictly speaking, the $g_5$ and $g_6$ terms cannot be absorbed into $g_1$, 
though their effects (relative to $g_1$) are expected to be rather small, due to the 
additional $\Lambda^2$ suppression.

It appears that a general analysis of the couplings of the Higgs boson candidate
would have to include at the very minimum the terms identified in eqs.~(\ref{fH2}-\ref{fA4}),  
and perhaps even the higher dimensional operators $\sim\Lambda^{-4},\Lambda^{-5},\ldots$
which were not explicitly listed. 
Such an analysis may indeed be desirable at some point in the future,
when significantly more data will have been accumulated by the
LHC experiments. 
However, there are strong theoretical and experimental motivations for
making certain simplifying assumptions for analyses in the \emph{immediate} future.

First, all experimental evidence so far suggests that the new physics scale 
$\Lambda$ is high compared to $v$. Second, consistency of the effective theory 
description {\em requires} that $\Lambda$ be sufficiently removed from the 
relevant experimental energy scale. Finally, the higher dimensional nature of the 
$1/\Lambda$ couplings suggests a radiative origin and hence suppressions
by loop factors. It is therefore reasonable to expect that the higher order terms 
$g_3$, $g_5$, $g_6$, etc.~in the 
expansions (\ref{fH2}-\ref{fA4}) are negligible relative to the corresponding
leading order couplings $g_1$, $g_2$ and $g_4$.

At the same time, the relative size of the leading terms $g_1$, $g_2$ and $g_4$
is {\em a priori} unknown. For example, the $g_2$ and $g_4$ terms are equally suppressed by
$\Lambda$, and may both preserve gauge invariance, thus it is difficult
to argue that one should be preferred over the other.
Similarly, we do not know the extent to which $H$ is involved in 
breaking $SU(2)$ gauge invariance, and hence we cannot simply assume that
the $SU(2)$ breaking term $g_1$ dominates over $g_2$, in spite of the
additional $\Lambda$ suppression in the latter \cite{impost}.
This is why we keep all three terms and consider the effective Lagrangian for
the mass eigenstate $X$ to be
\beq
\mathcal{L} = -X
\left[ \kappa_1 \frac{M_Z^2}{v} Z_\mu Z^\mu +
\frac{\kappa_2}{2 v} F_{\mu\nu} F^{\mu\nu} +
\frac{\kappa_3}{2 v} F_{\mu\nu}\tilde{F}^{\mu\nu}
\right],
\label{lagrangian}
\eeq
where $\tilde{F}_{\mu\nu} = \frac{1}{2} \varepsilon_{\mu\nu\rho\sigma}
F^{\rho\sigma}$ and
\beq
\kappa_1 \equiv g_1 \cos\alpha,
~
\kappa_2 \equiv g_2 \frac{v}{\Lambda} \cos\alpha,
~
\kappa_3 \equiv g_4 \frac{v}{\Lambda} \sin\alpha
\label{kappas}
\eeq
are the effective couplings which need to be measured by experiment.
We note that this procedure means neglecting (for the reasons stated above)
the $g_3$ operator from 
Refs.~\cite{jhu mem, mekd} and the
$g_5$ and $g_6$ terms listed in (\ref{fH2}).

%%%%%%%%%%%%% Begin OF FIGURE ################%%%%%%%%%%%%
\begin{figure}[t]
\includegraphics[width=8.0cm]{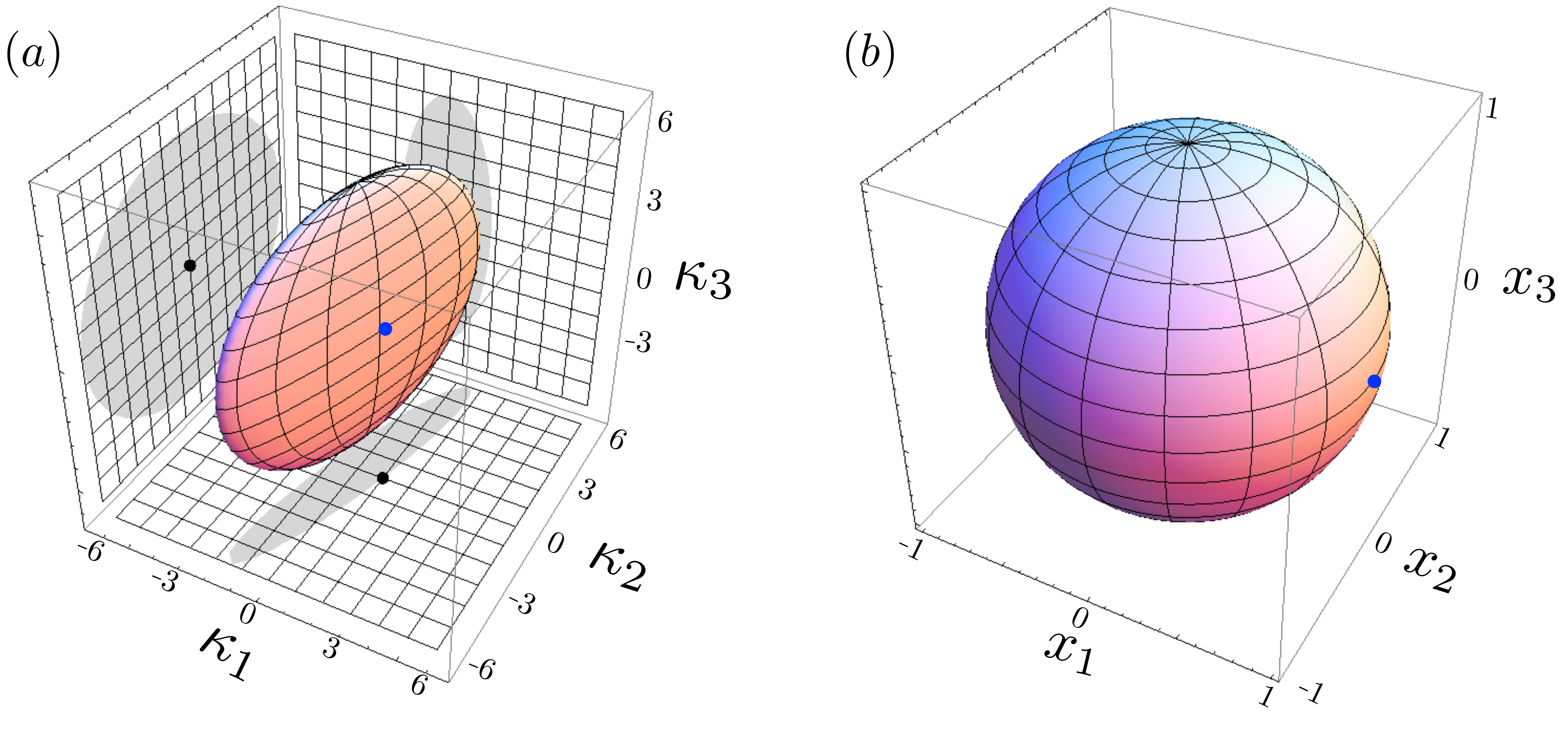} % Here is how to import EPS art
\caption{\label{fig:3Dplot}
The principle of geolocating: the change of variables (\ref{COV}) 
inflates (a) the hyper-surface defined by the measurement (\ref{GammaZZ})
into (b) a perfect sphere.}
\end{figure}
%%%%%%%%%%%%% End OF FIGURE %%%%%%%%%%%%%%%%%%%%%%%%%%%%%%%%%

{\bf Counting degrees of freedom.}  From an effective theory point of view,
the coefficients $\kappa_i$ in the Lagrangian (\ref{lagrangian}) are real numbers. 
Thus the problem of measuring the Higgs couplings is reduced to
determining the three dimensionless degrees of freedom $\kappa_i$.
However, one can go one step further and use the measured signal rate 
($X$ production cross-section times branching fraction for $X\to ZZ$) 
to eliminate one more degree of freedom, namely the overall $\kappa$ scale.
In order to do this, we will assume that the $X$ can be treated as a
narrow resonance.  We then note that the partial width for $X\to ZZ$
is a quadratic function of the $\kappa$'s:
\beq
\Gamma(X\to ZZ) = \Gamma_{SM} \sum_{i,j}\gamma_{ij}\kappa_i\kappa_j,
\label{GammaZZ}
\eeq
where we have factored out the partial width $\Gamma_{SM}$ predicted in the SM 
in order to define dimensionless coefficients $\gamma_{ij}$ listed in Table~\ref{gamma table}.
The measured total rate then provides the overall normalization and
constrains the $\kappa_i$ couplings to lie on the closed
hyper-surface shown in Fig.~\ref{fig:3Dplot}(a). The idea now is to
change variables and parameterize the couplings $\kappa_i$ in terms of the 
{\em two} coordinates on this hyper-surface. This re-parametrization is useful
and meaningful, since the (normalized) angular and invariant mass distributions used to measure the 
Higgs spin and CP properties are insensitive to the overall scale of the couplings $\kappa_i$. 
\begin{table}
\caption{\label{gamma table} Numerical values for the coefficients
defined in (\ref{GammaZZ}) for different diboson final states.}
\begin{ruledtabular}
\begin{tabular}{lcccc}
Process&$\gamma_{11}$&$\gamma_{22}$&$\gamma_{33}$&$\gamma_{12}$
\\
\hline
$X \to ZZ$ (DF)
& $1$ & $0.090$ & $0.038$ & $-0.250$ \\
$X \to ZZ$ (SF)
&$1$    & $0.081$ & $0.032$ & $-0.243$  \\
$X \to \gamma\gamma$
& $0$ & $1$ & $1$ & $0$ \\
$X \to WW$
& $1$ & $0.202$ & $0.084$  & $-0.379$  \\
\hline
\multicolumn{5}{c}{after cuts} \\
\hline
$X \to ZZ$ (DF)
&$1$  & $0.101$ & $0.037$ & $-0.277$
\\\end{tabular}
\end{ruledtabular}
\end{table}
Operationally, we propose to do this by changing variables as
\beq
x_i = \sum_{j} O_{ij} \kappa_j,
\label{COV}
\eeq
where $O_{21}= O_{31}= O_{32}=0$ and
\bea
O_{1i} &=& \gamma_{1i}/\sqrt{\gamma_{11}},~~ (i=1,2,3), \nonumber \\
O_{2i} &=& \frac{\gamma_{11}\gamma_{2i}-\gamma_{12}\gamma_{1i}}
            {\sqrt{(\gamma_{11}\gamma_{22}-\gamma_{12}^2)\gamma_{11}}},~~ (i=2,3),  \nonumber \\
O_{33} &=& \sqrt{\det ||\gamma_{ij}|| / (\gamma_{11}\gamma_{22}-\gamma_{12}^2)}. \nonumber
\eea
In terms of the new variables $x_i$, the constraint implied by (\ref{GammaZZ})
is the surface of a sphere, as seen in Fig.~\ref{fig:3Dplot}(b).

Note that in the case of real couplings, $\gamma_{13}$ and
$\gamma_{23}$ vanish identically (and hence are not listed
in~(\ref{GammaZZ})).
This can be understood in terms of parity; interference terms between
amplitudes describing the decay of a parity-even scalar and a
parity-odd scalar are odd under parity, hence vanish under integration
provided the cuts respect parity (as is nearly always the case).  The
situation in the presence of cuts is less clear if the couplings of
the three operators in~(\ref{lagrangian}) are complex valued; however, any
$\gamma_{13}$ or $\gamma_{23}$ term generated in this case is expected
to be very small.

\begin{figure}[t]
\includegraphics[width=4.2cm]{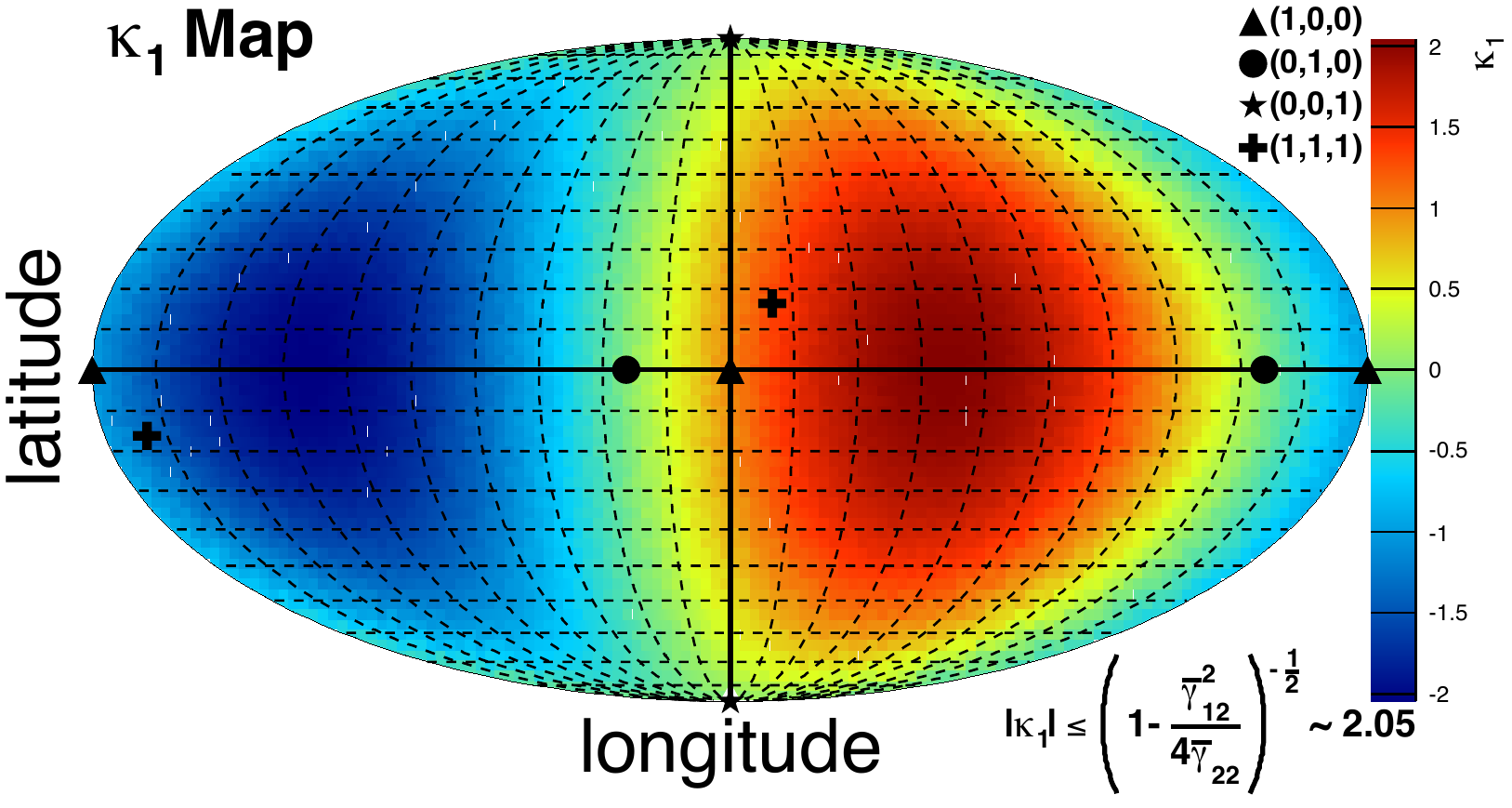} % Here is how to import EPS art
\includegraphics[width=4.2cm]{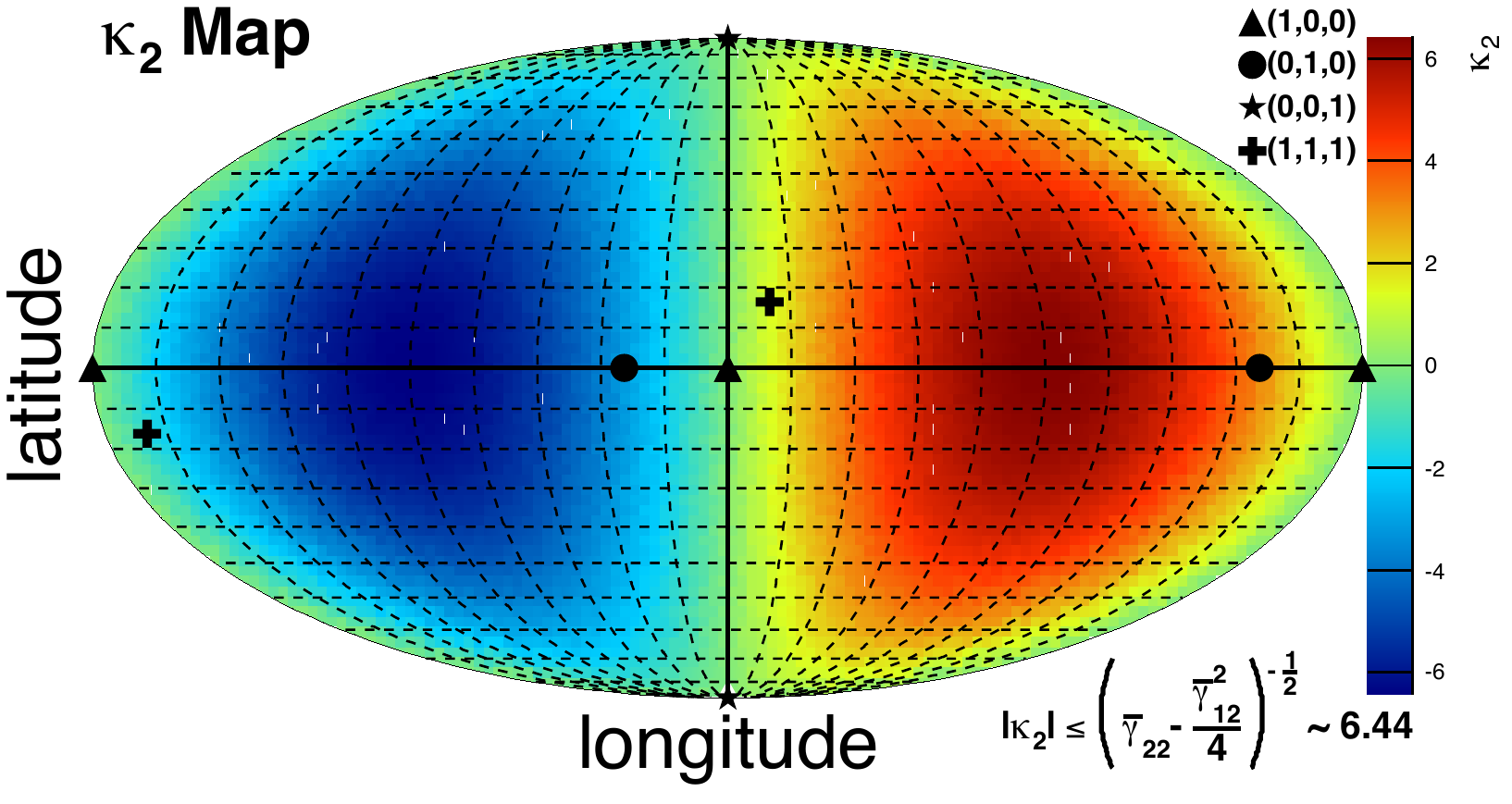} % Here is how to import EPS art
\includegraphics[width=4.2cm]{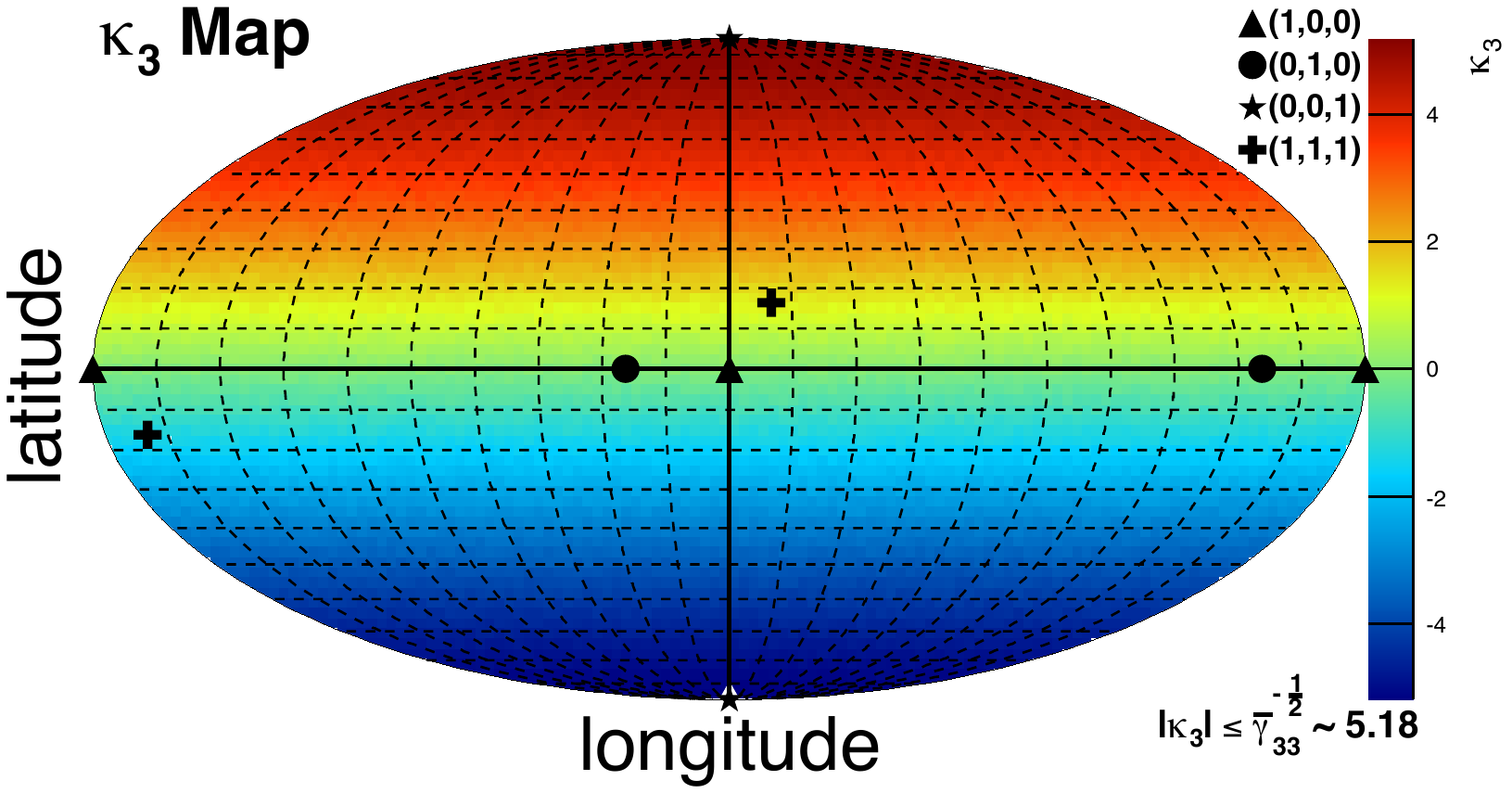} % Here is how to import EPS art
\includegraphics[width=4.2cm]{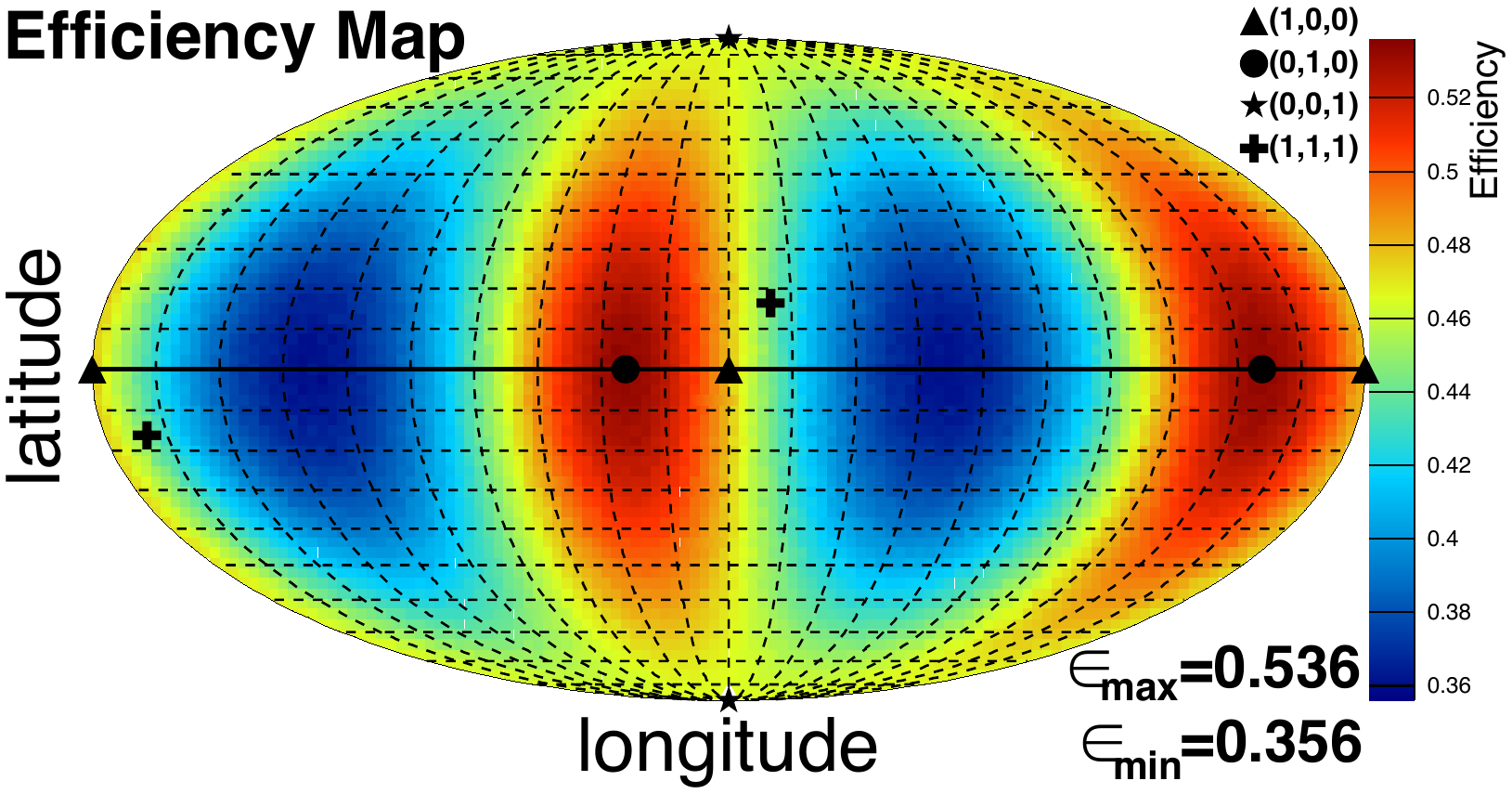} % Here is how to import EPS art
\caption{\label{fig:kappas}
The effective couplings $\kappa_i$ and the signal efficiency after cuts 
as a function of latitude ($\phi$) and longitude ($\lambda$) for the 
different flavor 4 lepton channel $X\to ZZ\to 2e2\mu$. }
\end{figure}

{\bf Geolocating the Higgs.} A sphere may be parameterized using
latitude $\phi$ and longitude $\lambda$.
Making an obvious analogy, we can think of the tree-level SM Higgs couplings 
$(\kappa_1,\kappa_2,\kappa_3)=(1,0,0)$ 
as being represented by the point $(\phi,\lambda)=(0,0)$ in the Gulf of Guinea
(see Fig.~\ref{fig:kappas}). The case of a pure pseudoscalar 
with $(\kappa_1,\kappa_2,\kappa_3)=(0,0,1)$ corresponds to 
the North pole with $(\phi,\lambda)=(90,0)$. 
Those are the two main benchmark scenarios 
considered so far by the LHC experiments. The Higgs sphere
from Fig.~\ref{fig:3Dplot}(b) now opens up the full range of possibilities.
For example, one may consider a certain amount of mixing as in (\ref{Xdef})
between the SM Higgs and a pseudoscalar, placing us on
the Greenwich meridian. Alternatively, one may allow non-trivial values for the two
$CP$-even couplings $\kappa_1$ and $\kappa_2$, thus spanning the equator.
Finally, one could also envision the most general case with non-trivial
values for all three couplings, 
e.g.~$(\kappa_1, \kappa_2, \kappa_3)=
(-0.945804, -3.88525, 2.44522)$,
corresponding to $(\phi,\lambda)=(29.64945,-82.3486)$, 
which happens to be a location in 
the south end zone of the Swamp.  
%  Finally, one could also envision the most general case with non-trivial
% values for all three couplings, e.g.~$(\kappa_1, \kappa_2, \kappa_3)=
% (-0.946, -3.885, 2.445)$,
% corresponding to $(\phi,\lambda)=(29.65,-82.35)$.

%%%%%%%%%%%%% Begin OF FIGURE ################%%%%%%%%%%%%
%%%%%%%%%%%%% End OF FIGURE %%%%%%%%%%%%%%%%%%%%%%%%%%%%%%%%%

We note that CMS has recently undertaken \cite{CMS-PAS} an exploration of the 
sphere along the Greenwich meridian ($\lambda=0$) by measuring
the parameter $f_{a3}\equiv |A_3|^2/(|A_3|^2+|A_1|^2)$, 
parametrizing the relative contributions $|A_1|^2$ and $|A_3|^2$
to the total cross-section of the two benchmark models 
$(\kappa_1, \kappa_2, \kappa_3)=(1,0,0)$ and 
$(\kappa_1, \kappa_2, \kappa_3)=(0,0,1)$, respectively.

\begin{figure}[t]
\includegraphics[width=5.0cm]{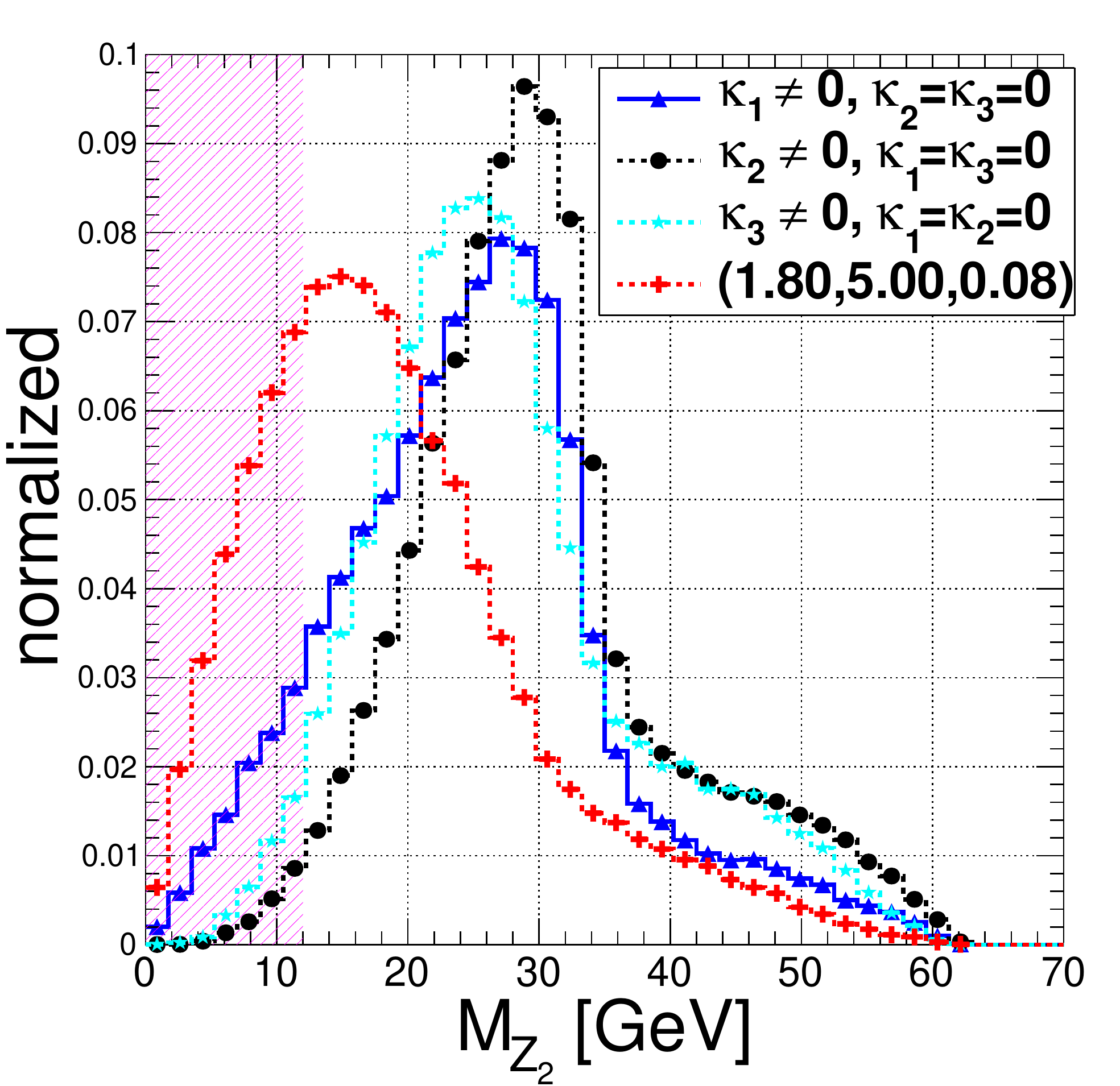} % Here is how to import EPS art
\caption{\label{fig:MZ2}
Unit-normalized $M_{Z_2}$ distributions for four different theory benchmarks
$(\kappa_1,\kappa_2,\kappa_3)$. }
\end{figure}

{\bf Selection cut bias.} By definition, the measurement of the Higgs signal rate
(and, by association, of the partial width (\ref{GammaZZ})) is done using only 
events which pass selection cuts. Therefore, it needs to be corrected for the efficiency.
Unfortunately, as demonstrated in the bottom right panel in Fig.~\ref{fig:kappas},
the signal efficiency is noticeably model dependent, i.e.~it is quite sensitive to the
actual values of the individual $\kappa_i$ parameters, and may vary from
as low as $35\%$ to as high as $54\%$. One source of this variation is 
the cut on $M_{Z_2}$ (see e.g.~\cite{hzz_m2}). The effect is quantified 
in Fig.~\ref{fig:MZ2}, which shows unit-normalized $M_{Z_2}$ distributions for
several choices of $\kappa_i$, including an example with a high efficiency 
($(\kappa_1,\kappa_2,\kappa_3)=(0,0,1)$, black circles) and a suitably chosen 
example with a low efficiency ($(\kappa_1,\kappa_2,\kappa_3)=(1.8,5,0.08)$, red crosses).
The $M_{Z_2}$ distribution in the latter case is softer, and many events fall below the minimum
accepted $M_{Z_2}$ value of 12 GeV.
We conclude that it would be virtually impossible to correct for the efficiency 
without knowledge of the couplings $\kappa_i$, which we are trying to measure 
in the first place, thus falling into a vicious circle.

%%%%%%%%%%%%% Begin OF FIGURE ################%%%%%%%%%%%%
%%%%%%%%%%%%% End OF FIGURE %%%%%%%%%%%%%%%%%%%%%%%%%%%%%%%%%

Here we advocate an alternative approach.
It is sufficient to realize that the coefficients $\gamma_{ij}$, while also
affected by the efficiency, do {\em not} depend on $\kappa_i$.
Therefore, in defining the Higgs sphere (\ref{COV}) one could simply use 
the corresponding values of $\gamma_{ij}$ {\em after cuts}, 
which need to be calculated once and for all.
As shown in Table~\ref{gamma table}, the changes are subtle, yet
noticeable. Note that even without any cuts, the same flavor (SF) and 
different flavor (DF) coefficients are different, due to the interference effects 
present in the SF case (see, e.g.~\cite{mekd}).

\begin{figure}[t]
\includegraphics[width=4.2cm]{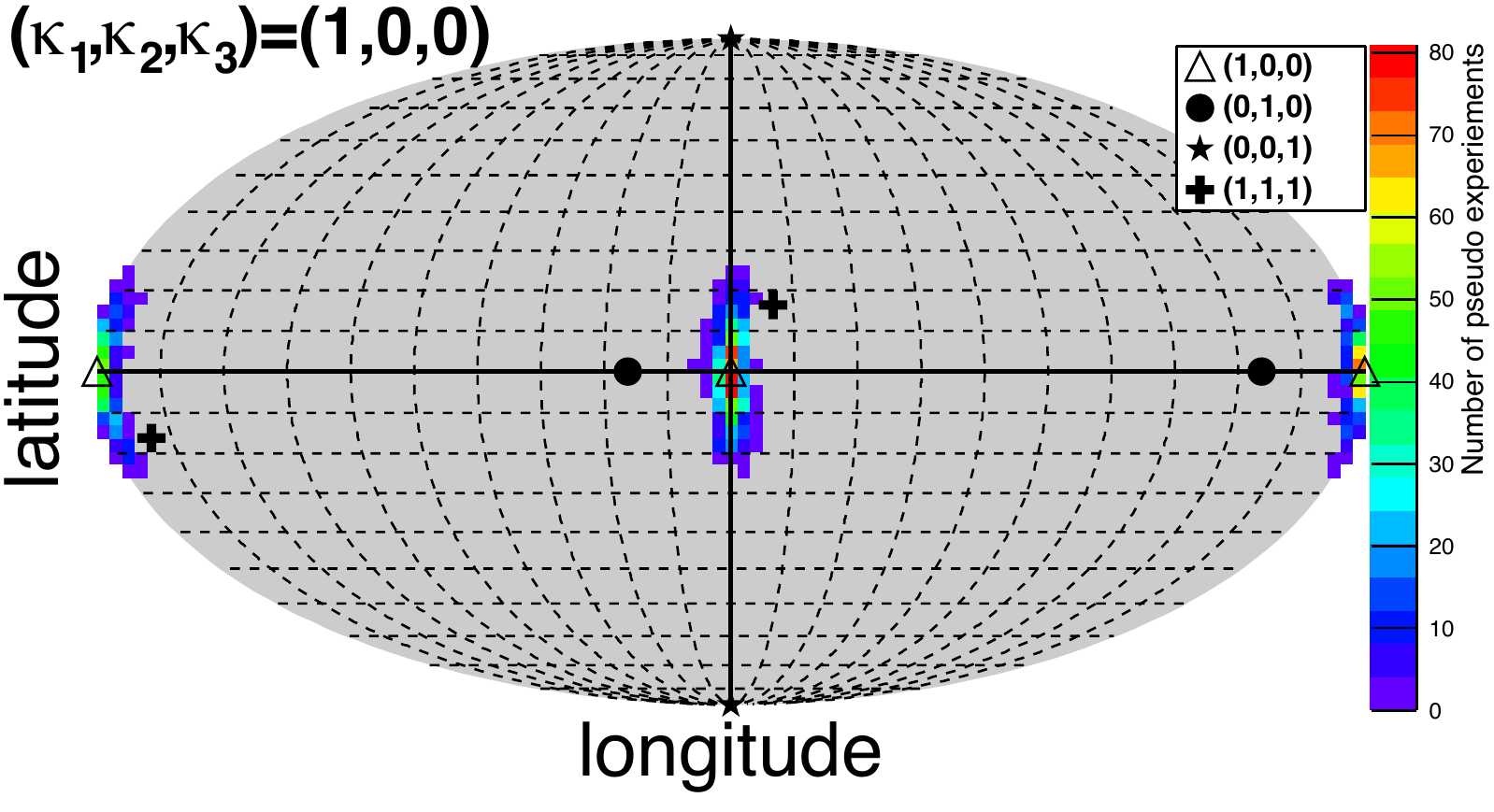} % Here is how to import EPS art
\includegraphics[width=4.2cm]{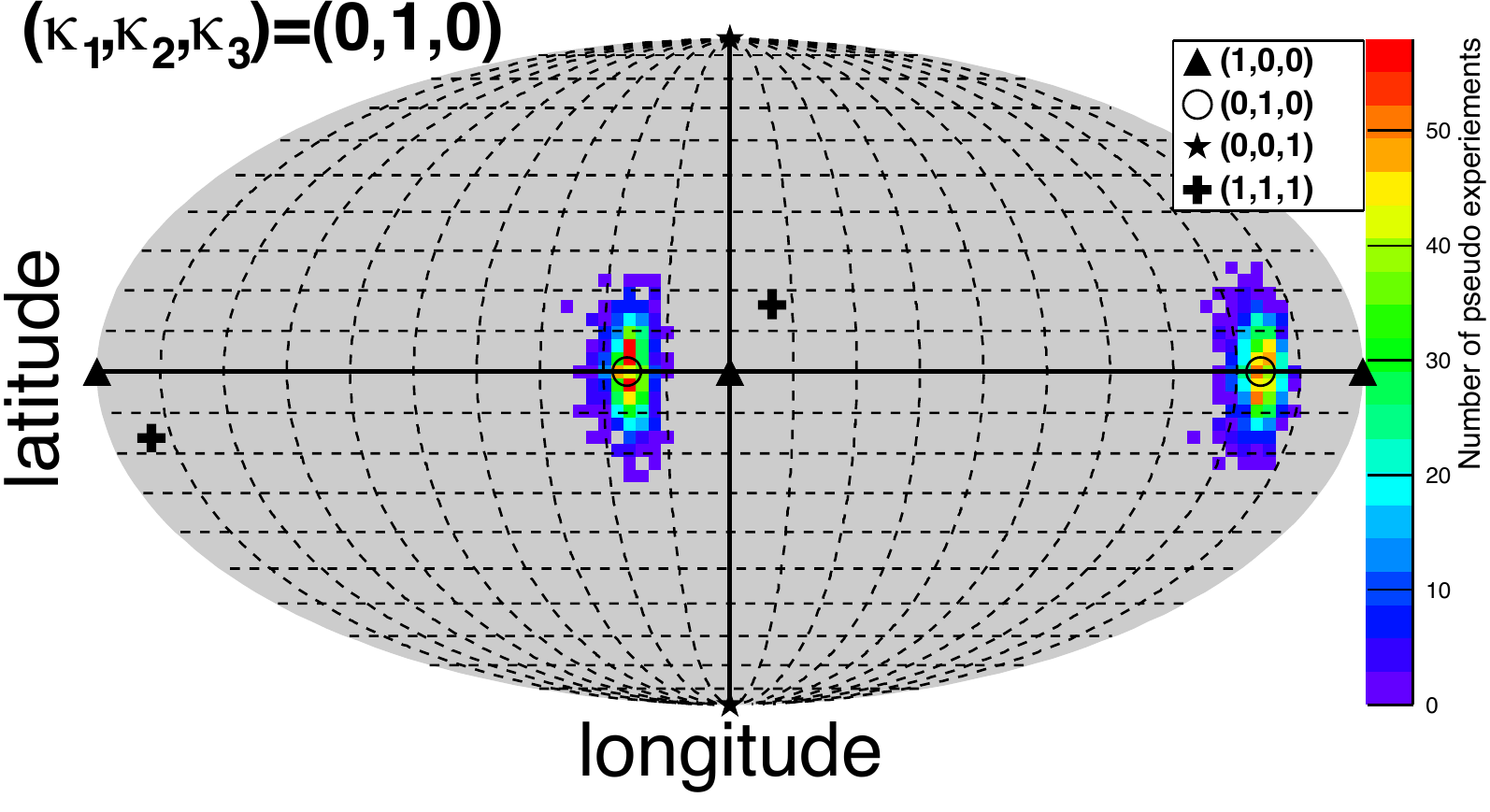} % Here is how to import EPS art
\includegraphics[width=4.2cm]{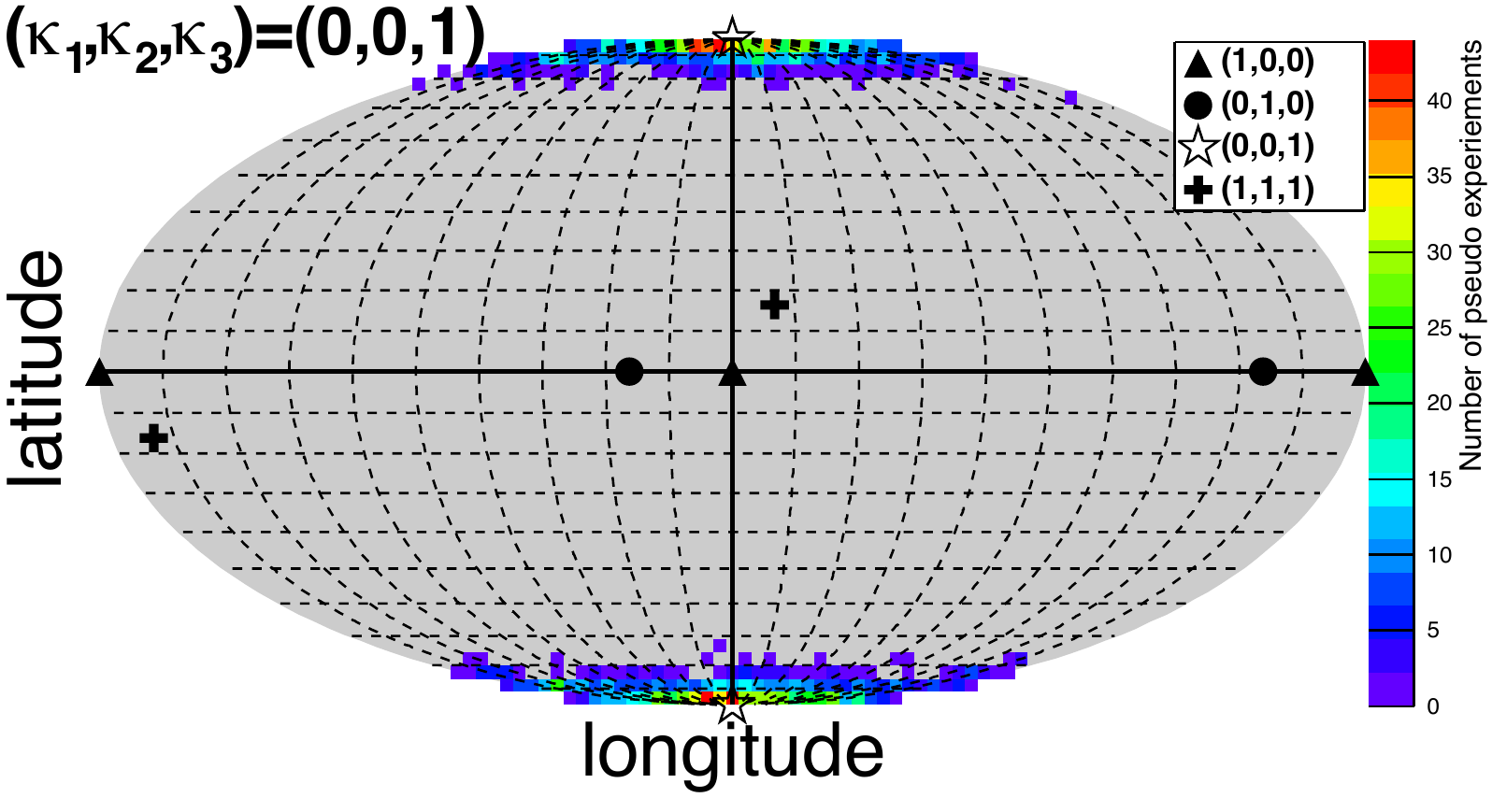} % Here is how to import EPS art
\includegraphics[width=4.2cm]{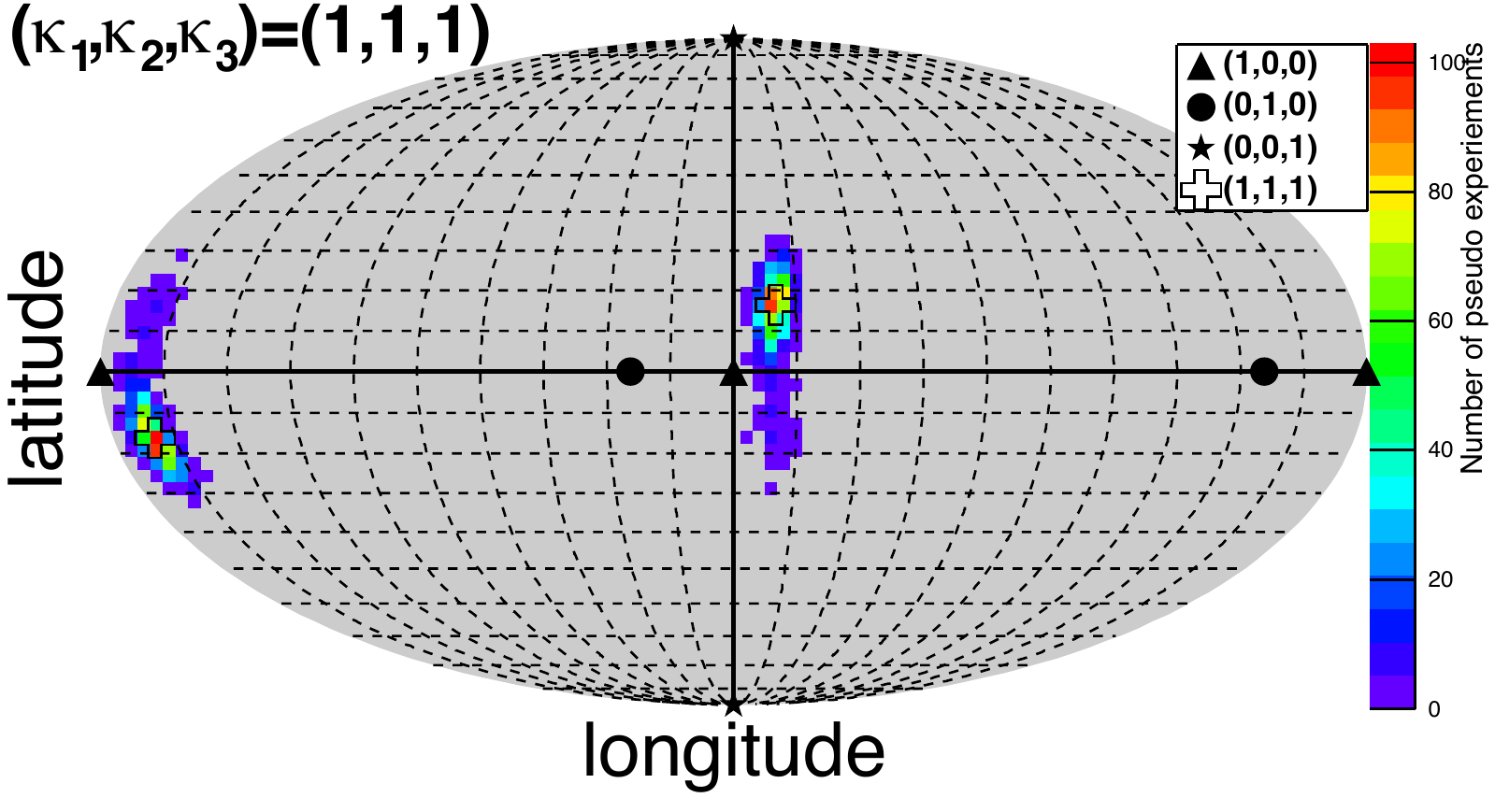} % Here is how to import EPS art
\caption{\label{fig:Sc1}
Distribution of the best fit values for the Higgs geolocation in 1000 pseudoexperiments
with 300 $X\to ZZ\to 2\mu2e$ signal events each, for four different benchmark points. }
\end{figure}

{\bf Interpretation.} We emphasize that by studying the angular and invariant 
mass distributions of the final state leptons in the $X\to ZZ\to 4\ell$ channel 
{\em it is experimentally possible} to determine the exact geolocation of the 
Higgs boson candidate, without any simplifying theoretical assumptions. 
A proof of principle is shown in Fig.~\ref{fig:Sc1}. We consider 4 benchmark
scenarios for $(\kappa_1,\kappa_2,\kappa_3)$: 
($\triangle$) $(1,0,0)$,
($\circ$) $(0,1,0)$,
($\star$) $(0,0,1)$, and 
($+$) $(1,1,1)$.
We show results from 1000 pseudoexperiments with 300 signal events each (after cuts).
Fig.~\ref{fig:Sc1} shows that in each case, the maximum likelihood fit indeed selects the 
correct geolocation (marked with an open symbol) of the Higgs candidate.

%%%%%%%%%%%%% Begin OF FIGURE ################%%%%%%%%%%%%
%%%%%%%%%%%%% End OF FIGURE %%%%%%%%%%%%%%%%%%%%%%%%%%%%%%%%%

{\bf (In)determination of the mixing angle $\alpha$.}
We note that geolocating the Higgs candidate by itself is not sufficient to determine the
value of the mixing angle $\alpha$, and we will still not know the relative 
composition of $X$ in terms of $H$ and $A$. 
As seen in eq.~(\ref{kappas}), the measurement of all three effective couplings
$\kappa_1$, $\kappa_2$ and $\kappa_3$ only places 3 constraints on the 4 input parameters
$\alpha$, $g_1$, $g_2/\Lambda$ and $g_4/\Lambda$, so
one degree of freedom (which can be chosen to be the 
mixing angle $\alpha$) will always be left undetermined.

{\bf Music of the Spheres.}
Until now, we have been treating the coefficients $\kappa_i$
of the operators in the effective Lagrangian (\ref{lagrangian}) to be real.
We term this ``Scenario~0''.
Of course, this must be the case at tree level, and even beyond,
as long as those operators are generated by loop diagrams
involving heavy particles, such as the top, which may be consistently
integrated out of the full theory to form an effective theory.
However, in accordance with the optical theorem, loops with light 
particles (such as $b$-quarks, which cannot be integrated out of the theory) 
will contribute imaginary parts to those couplings. 
In this case, each of the $\kappa_i$
is complex and there are five physical degrees of freedom
(an overall phase in the $\kappa_i$ may be removed).
Nevertheless, the width constraint 
$\Gamma(X\to ZZ) = \Gamma_{SM} \sum_{i,j}\gamma_{ij}\kappa_i^\ast\kappa_j$
can still be enforced. As before, the diagonalization and rescaling~(\ref{COV})
renders this a sphere in $\mathbb{C}^3$
(rather than an ellipsoid, as is the case in general).  

Even in this general complexified case, there are three interesting 
and simple scenarios where the number of degrees of
freedom is reduced to three.  Namely, if one of $\kappa_i$ is too
small to have observable effects, then the remaining two $\kappa_i$,
taken to be arbitrary complex numbers, also comprise two observable
degrees of freedom (as an overall phase may be removed).  There are,
of course three such scenarios, ``Scenario~1'' where $\kappa_1$ is
negligible, ``Scenario~2'' where $\kappa_2$ is negligible, and
``Scenario~3'' where $\kappa_3$ is negligible.

%All four scenarios have physical motivation.  Scenario~0 holds when
%the important couplings occur at tree-level or come largely from
%top-loops.
%Scenario~1 describes the most general couplings to a massless
%particle such as the photon or the gluon, though it could be obtained in
%the $ZZ$ or $WW$ cases if the $X$ has nothing to do with EWSB.
%Scenario~2 describes, e.g., a state where $X$ is primarily $A$, but
%the $H$ with which it mixes is an SM-like Higgs boson.  Scenario~3
%describes $X$ as a general CP-even scalar.  The scenarios are not
%totally exclusive; for instance, the $ZZ$ case
%in the Standard Model is well approximated by points in Scenarios~0,
%2, or 3.

{\bf Other final state channels.} While we have so far only focused on the 
$ZZ\to 4\ell$ final state, the preceding discussion can be readily applied to the other
di-boson final states, 
$W^+W^-$, %\cite{WWpapers}, 
$Z\gamma$, % \cite{ZGpapers},
and
$\gamma\gamma$. %\cite{GGpapers}. 
The interactions relevant for those channels 
involve corresponding new couplings $\kappa_i^{\tiny WW}$, 
$\kappa_i^{\tiny Z\gamma}$,  and $\kappa_i^{\tiny \gamma\gamma}$.
Assuming $SU(2)$ gauge symmetry, these new couplings are related, e.g.
\bea
\kappa_1^{\gamma\gamma}&=& \kappa_1^{Z\gamma}=0,~~~\kappa_1^{WW}= \kappa_1^{ZZ},  \\
\kappa_i^{Z\gamma} &=& \frac{1}{2}(\kappa_i^{\tiny ZZ}- \kappa_i^{\tiny \gamma\gamma})\tan2\theta_W, (i=2,3),\\
\kappa_i^{WW} &=&  \frac{\kappa_i^{\tiny ZZ}\cos^2\theta_W - \kappa_i^{\tiny \gamma\gamma}\sin^2\theta_W}{\cos^2\theta_W - \sin^2\theta_W}, (i=2,3),
\eea
with obvious definitions of $\kappa$'s.
As before, the measured rate in each of those channels defines a sphere similar to 
eq.~(\ref{GammaZZ}), where the corresponding values for the coefficients $\gamma_{ij}$
are given in Table~\ref{gamma table}. It is interesting to note that these spheres are not 
overlapping (even for the DF and SF case in the same $X\to ZZ$ channel), 
and in principle couplings can be measured from intersecting spheres,
although we expect that this method is of only academic interest, since 
the corresponding precision would be very poor.

{\bf Summary.}
We have proposed several related parameterizations of the couplings of
the $125$ GeV boson discovered at the LHC.  These parameterizations
allow the LHC experiments to go beyond comparing benchmark points,
etc. in a relatively simple, yet very general, theoretically motivated
fashion.  The
low dimensionality of the parameter space is helpful experimentally as
well as in the visualization and interpretation of results.
We look forward to the implementation of this method by the LHC
collaborations and to the continued exploration of the couplings of
the putative Higgs boson.

{\bf Acknowledgements.}
We thank A.~Gritsan, A.~Korytov, I.~Low, and G.~Mitselmakher for useful discussions.  
JG, JL, KM and SM thank their CMS colleagues.
JL acknowledges the hospitality of the SLAC Theoretical Physics Group.
MP is supported by the CERN-Korea fellowship through the National Research
Foundation of Korea.  Work supported in part by U.S. Department of
Energy Grants DE-FG02-97ER41029.  Fermilab is operated by the Fermi
Research Alliance under contract DE-AC02-07CH11359 with the
U.S. Department of Energy.

\end{document}